\documentclass[]{arxsub}

\usepackage{ifluatex}
\ifluatex
\usepackage{fontspec}
\usepackage{polyglossia}
\setdefaultlanguage{english}
\else
\usepackage[english]{babel}
\usepackage[utf8]{inputenc}
\fi

\usepackage[
  backend=biber,
  natbib = true,
  style=ext-numeric-comp,
  sorting=none,
  minbibnames=3, maxbibnames=6,
  uniquename=false, uniquelist=false,
  giveninits=true, terseinits,
  articlein=false, innamebeforetitle=true,
  isbn=false,
  urldate=long, dateabbrev=false,
  autocite=superscript,
]{biblatex}

\DeclareNameAlias{sortname}{family-given}

\DeclareNameAlias{author}{sortname}
\DeclareNameAlias{editor}{sortname}
\DeclareNameAlias{translator}{sortname}

\DeclareDelimAlias{finalnamedelim}{multinamedelim}

\DeclareFieldFormat
  [article,inbook,incollection,inproceedings,patent,thesis,unpublished]
  {title}{#1}

\renewbibmacro*{journal+issuetitle}{\usebibmacro{journal}\setunit*{\addspace}\iffieldundef{series}
    {}
    {\newunit
     \printfield{series}\setunit{\addspace}}\usebibmacro{issue+date}\setunit{\addsemicolon\addspace}\usebibmacro{volume+number+eid}\setunit{\addcolon\space}\usebibmacro{issue}}

\DeclareFieldFormat[article,periodical]{number}{\mkbibparens{#1}}

\renewbibmacro*{issue+date}{\usebibmacro{date}}

\renewbibmacro*{pubinstorg+location+date}[1]{\printlist{location}\iflistundef{#1}
    {\setunit*{\addsemicolon\space}}
    {\setunit*{\addcolon\space}}\printlist{#1}\setunit*{\addsemicolon\space}\usebibmacro{date}\newunit}

\DeclareFieldFormat{pages}{#1}

\DeclareFieldFormat{doi}{doi\addcolon\space
  \ifhyperref
    {\href{https://doi.org/#1}{\nolinkurl{#1}}}
    {\nolinkurl{#1}}}

\AtEveryBibitem{\clearfield{number}\clearfield{month}\clearfield{day}\clearfield{urlyear}\clearfield{urlmonth}\clearfield{issn}\clearfield{series}\clearfield{pagetotal}\clearfield{eprintclass}\ifentrytype{article}{
	\iffieldundef{volume}{}{\clearfield{doi}}\clearfield{eprint}\clearfield{url}}{}\ifentrytype{book}{\clearfield{eprint}\clearfield{doi}\clearfield{url}}{\clearname{editor}}
\ifentrytype{inproceedings}{\clearfield{eprint}\clearfield{url}\clearfield{doi}\clearfield{publisher}\iffieldundef{booktitle}{}{\clearfield{eventtitle}}}{}\iffieldundef{eprint}{}{\clearfield{url}}
\iffieldundef{doi}{}{\clearfield{url}}
}

\usepackage{csquotes}
\usepackage{siunitx}
\DeclareSIUnit\pixel{pixel}
\usepackage[section]{placeins}

\usepackage{hyperref}
\usepackage[capitalise,noabbrev]{cleveref}

\usepackage{listings}
\usepackage{booktabs}

\providecommand{\doi}{}
\renewcommand{\doi}[1]{\href{https://doi.org/#1}{#1}}

\newcommand{\preSubmission}{Part of this work has been presented at the ISMRM Annual Conference 2026 (Cape Town).}

\providecommand{\del}[2][]{}
\providecommand{\dels}[2][]{}

\makeatletter
\appto{\appendix}{\setcounter{figure}{0}}
\makeatother

\begin{document}

\articletype{Toolbox Article}
\title{BART Online Open-Source Sequence Toolbox for Computational MRI}

\corres{Martin Uecker, Graz University of Technology,
Institute of Biomedical Imaging, Stremayrgasse~16/3, 8010~Graz, Austria, \email{uecker@tugraz.at}}

\author[1]{Daniel Mackner}{}
\author[1]{Philip Schaten}{}
\author[1,2,3]{Markus Huemer}{}
\author[1]{Viktoria Buchegger}{}
\author[1,3]{Moritz Blumenthal}{}
\author[3]{Xiaoqing Wang}{}
\author[1,4,5]{Martin Uecker}{}

\authormark{D. Mackner \textsc{et al.}}

\address[1]{Institute of Biomedical Imaging, Graz University of Technology, Graz, Austria}
\address[2]{Athinoula A. Martinos Center for Biomedical Imaging, Department of Radiology, Massachusetts General Hospital, Charlestown, USA}
\address[3]{Department of Radiology, Harvard Medical School, Boston, Massachusetts, USA}
\address[4]{German Centre for Cardiovascular Research (DZHK), Partner Site Göttingen, Göttingen, Germany}
\address[5]{BioTechMed-Graz, Graz, Austria}

\finfo{
This work has been supported by the DZHK (German Centre for Cardiovascular
Research), funding code: 81Z0300115. This research was funded in part by NIH
under grant U24EB029240. This research was funded in whole or in part 
by the Austrian Science Fund (FWF) 10.55776/F100800.}
\newpage

\abstract{
\section{Purpose}
In advanced computational MRI techniques, acquisition and reconstruction techniques
are jointly designed. For reproducibility, it is therefore important to 
provide an open implementation of both. At the same time, any use in a clinical
environment usually requires a close integration with the MRI scanner. Ensuring 
long-time reproducibility and maintenance then poses additional challenges.
In this work, we aim  to provide a fully integrated open-source framework that
can meet these demands.

\section{Methods}
A software framework to develop pulse sequences is added to BART,
an open-source toolbox for computational MRI.
In addition, a vendor-specific driver sequence is developed 
that can be used to run the sequence on a clinical MRI scanner enabling
online adjustment of all relevant sequence parameters.  Using the
Pulseq format, the exact same sequence can also be reproduced offline
using a widely used vendor-neutral open-source standard.
As proof-of-concept, quantitative MRI  methods for $T_1$ and joint water/fat 
$R_2^*,~B_0$ mapping using radial FLASH  and model-based reconstruction
are implemented in the proposed framework. Consistency  between online
and offline acquisition is validated in phantom and in vivo experiments.

\section{Results}
Quantitative MRI methods highlighting specific challenges
of acquisition and reconstruction
were successfully implemented in BART. Acquisition parameters and FOV
can be adapted online on a clinical MRI system. Quantitative parameter
maps from model-based reconstruction agree for online and offline
regenerated Pulseq acquisitions.

\section{Conclusion}
This work enables reproducibility of advanced computational
MRI methods within a comprehensive end-to-end open-source framework.
}


\jnlcitation{\cname{\author{D. Mackner},
\author{P. Schaten},
\author{M. Huemer},
\author{V. Buchegger},
\author{M. Blumenthal},
\author{X. Wang},
\author{M. Uecker}} (\cyear{2026}),
\ctitle{BART Sequence},
\cjournal{Magn. Reson. Med.},
\cvol{???}.}

\maketitle

\section{Introduction}
\label{sec:intro}

Over the last decades, many new MRI techniques were developed to achieve 
faster acquisitions, higher resolutions, and to enable entirely new applications,
especially for quantitative MRI (qMRI).  With increased complexity of both, 
pulse sequences and reconstruction algorithms, reproducibility of 
published methods requires substantial effort due to missing details.
For this reason, reproducibility  became an important topic
within the MRI community \cite{Stikov_Magn.Reson.Med._2019,
Knoll_Radiol.Artif.Intell._2020, Veldmann_Magn.Reson.Med_2022,Tamir_MAGMA_2025}.

Pulse sequences are typically implemented using a vendor-specific proprietary
platform, hampering translation across systems and sharing of source code
within the research community. To overcome these limitations, several standards 
and frameworks for sequence development have been developed
\cite{Jochimsen_J.Magn.Reson._2004,Magland_Magn.Reson.Med._2016,Nielsen_Magn.Reson.Med._2018}.
While many frameworks did not find traction in the community so far,
Pulseq \cite{Layton_Magn.Reson.Med._2017} together with pypulseq
\cite{Ravi_Magn.Reson.Imaging_2018,Ravi_J.OpenSourceSoftw._2019},
became a vendor-neutral open-source standard for sharing MRI sequences.
Numerous applications have been published using the format,
including MR fingerprinting \cite{Schuenke_MAGMA_2025,Griesler_Magn.Reson.Med._2026}
cardiac $T_1$ mapping \cite{Gaspar_Magn.Reson.Med._2024,Gaspar_Magn.Reson.Med._2023},
and Chemical Exchange Saturation Transfer (CEST) \cite{Herz_Magn.Reson.Med._2021}.
Pulseq files are human-readable \verb|.seq|-files that can be executed on  MRI scanners 
of different vendors using vendor-specific driver sequences and can be
used as input for Bloch simulations
\cite{Loktyushin_Magn.Reson.Med._2021,Castillo-Passi_Magn.Reson.Med._2023,
Weinmueller_Magn.Reson.Med._2025}.
Environments for open-source pulse sequence development now usually
offer the possibility to convert to Pulseq \cite{Stoecker_Magn.Reson.Med._2010,Artiges_Magn.Reson.Med._2026}.
While highly flexible in the sequence design and fully reproducible, 
the Pulseq format represents a pre-computed MRI sequence with limited
possibilities to adapt the scan to each patient, which hampers its use 
in clinical studies.
gammaSTAR \cite{Konstandin_Magn.Reson.Med._2025} is a vendor-agnostic
framework for pulse sequences that allows online adjustment of sequence
parameters, but it is not open source.

In this work, we describe an open-source sequence programming framework,
which is developed as an extension of the Berkeley Advanced Reconstruction
Toolbox (BART) \cite{Uecker__2013}.
BART is an open-source toolbox for computational MRI. It
includes algorithms for
iterative optimization, efficient calibration and image
reconstruction for parallel imaging and compressed sensing,
advanced model-based reconstruction for quantitative MRI,
as well as for machine learning.
In addition to this basic framework, we developed a driver sequence 
that allows online configuration of the sequence on the MRI scanner.
Sequences can also be exported to Pulseq files and later reproduced exactly.

To validate feasibility of the framework for development of advanced
computational MRI techniques, we implemented two quantitative methods
that combine radial acquisition with
model-based reconstruction \cite{Fessler_IEEETrans.SignalProcessing_2003,Uecker__2013,
Graff__2006,Olafsson_IEEETrans.Med.Imag._2008,
Block_IEEETrans.Med.Imaging_2009,Sumpf_J.Magn.Reson.Imaging_2011,
Wang_Philos.Trans.R.Soc.A._2021,maier2020pyqmri,Zimmermann__2026}.  
In these methods, the model-based reconstruction is sensitive to the
specific details of the acquisition such as k-space trajectory and 
timing ($T_I$, $T_R$, $T_E$), highlighting the need for an integrated
approach.
Reproducibility of these techniques is validated in phantom and 
in vivo experiments by comparing the results when using the online driver
sequence with results obtained when reproducing the acquisition 
using exported Pulseq files.

\newpage
\section{Architecture and Implementation}
\label{sec:arch}

\begin{figure}[h!]
\centering
\includegraphics[width=0.85\textwidth]{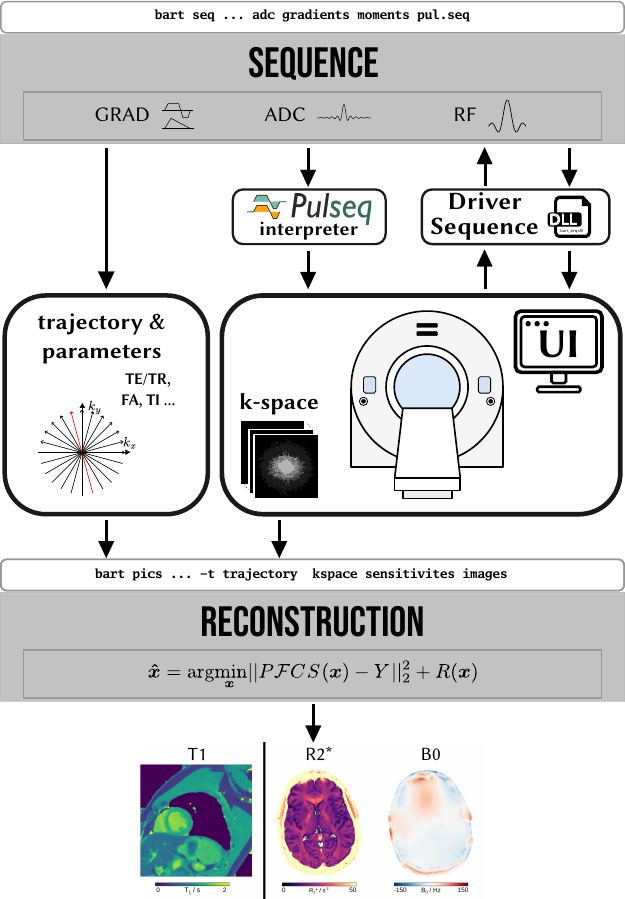}
\caption{Overview of the proposed open-source computational MRI framework.
It is implemented in BART with both, online integration with a
driver sequence  and offline generation of Pulseq files with
the BART \texttt{seq} command.
Sequences are uniquely described as a set of events, which are
interpreted during the acquisition. Parameters required for image
reconstruction, e.g. the trajectory, can be derived from the events. 
Integration of acquisition and reconstruction into a consistent framework
helps to ensure consistencybetween both in model-based qMRI.}
\label{fig:seq}
\end{figure}

A schematic overview of the BART sequence framework can be found in
Figure \ref{fig:seq}.

\subsection{Sequence Definitions}

The specific configuration of an MRI sequence depends on a 
high-level parameter set, comprising fixed system parameters as 
well as adjustable parameters. System parameters include the maximum gradient strength, slew rate, 
or raster times. Adjustable parameters include geometric information
(FOV, base resolution), the typical timing and contrast-related settings ($T_E$/$T_R$, flip angle),
magnetic preparations (inversion pulse, $T_I$, delays), as well
as high-level dimensions (spokes, slices) and phase encoding settings.
Similar to other sequence development frameworks, we compose MRI sequences from 
different event types, namely gradient events, RF pulses, 
sampling (Analog-to-digital converter, ADC) periods, wait and trigger events, which we each
parametrize by start-, mid- and endtime, and additional type specific parameters.
\paragraph{Gradients} are decomposed into triangles,
each described by its amplitude at midtime in phase-, read- and slice-encoding direction.
Hence, typical trapezoidal gradients have a compact representation using
two overlapping asymmetrical triangles, while
arbitrary gradient shapes can be designed by superposition of 
multiple gradient events corresponding to a linear spline interpolation.
The MRI scanner requires that the gradient is computed on a discrete raster
. This is done by integrating the linear splines over each system-specific
gradient raster interval and divided by this raster time to obtain the average gradient
for each interval. 
When assuming Dirac pulses or boxcar functions as representation of 
the discretized gradient waveform, using these averages as coefficients
then ensures that the zeroth moments of the continuous and discretized
waveforms are identical at the end of each raster interval. Magnitude and
slew rate constraints are checked on the original continuous waveform and
these constraints are then automatically fulfilled also for the 
computed discrete amplitudes and for finite differences, respectively.
\paragraph{RF events} describe a pulse
with flip angle, phase, a frequency shift, and a mapping to a pulse shape.
Pulse shapes are prepared prior to sequence
execution. This preparation includes SAR safety checks (with duration, shape and
number of pulse applications) where parameters can be adapted using the vendor-specific
solve handler if limits are exceeded. In addition, we perform feasibility checks for
the applied RF amplitudes, where we ensure that the amplitude is not reduced
due to hardware limitations.
A frequency offset and phase can be configured for the event, where the
phase is defined for the midtime, with respect to the symmetry of the
pulse shape. Non-linear phases (e.g. for adiabatic pulses) are described with
complex-valued RF samples.
Pulse events are prepared on demand based on a mapping to previously prepared 
shapes.
\paragraph{ADC events} are defined with a
dwell-time and number of samples, as well as frequency and phase information,
again using the midtime as a reference for the phase.
ADC samples are located at the center of the dwell time intervals.
In addition, each ADC event includes a vector of loop indices which are 
then stored together with the acquired data.
\paragraph{Trigger/wait events} can be used to synchronize the measurement
with external devices, i.e. by waiting for a specific condition 
or by sending a signal to an external device. Wait events can be
used to insert delays. Parameters for those events are extracted from
the respective timing information.

A set of events is combined into a sequence block, which are classified
into magnetic preparation or imaging blocks. All events in a block are
prepared as a function of the sequence parameters and the current
sequence state. This state depends on multi-dimensional loop
indices, which may indicate the current slice, spoke, or frame.

\subsection{Online Sequence}
\begin{figure}
	\centering
	\includegraphics[width=\textwidth]{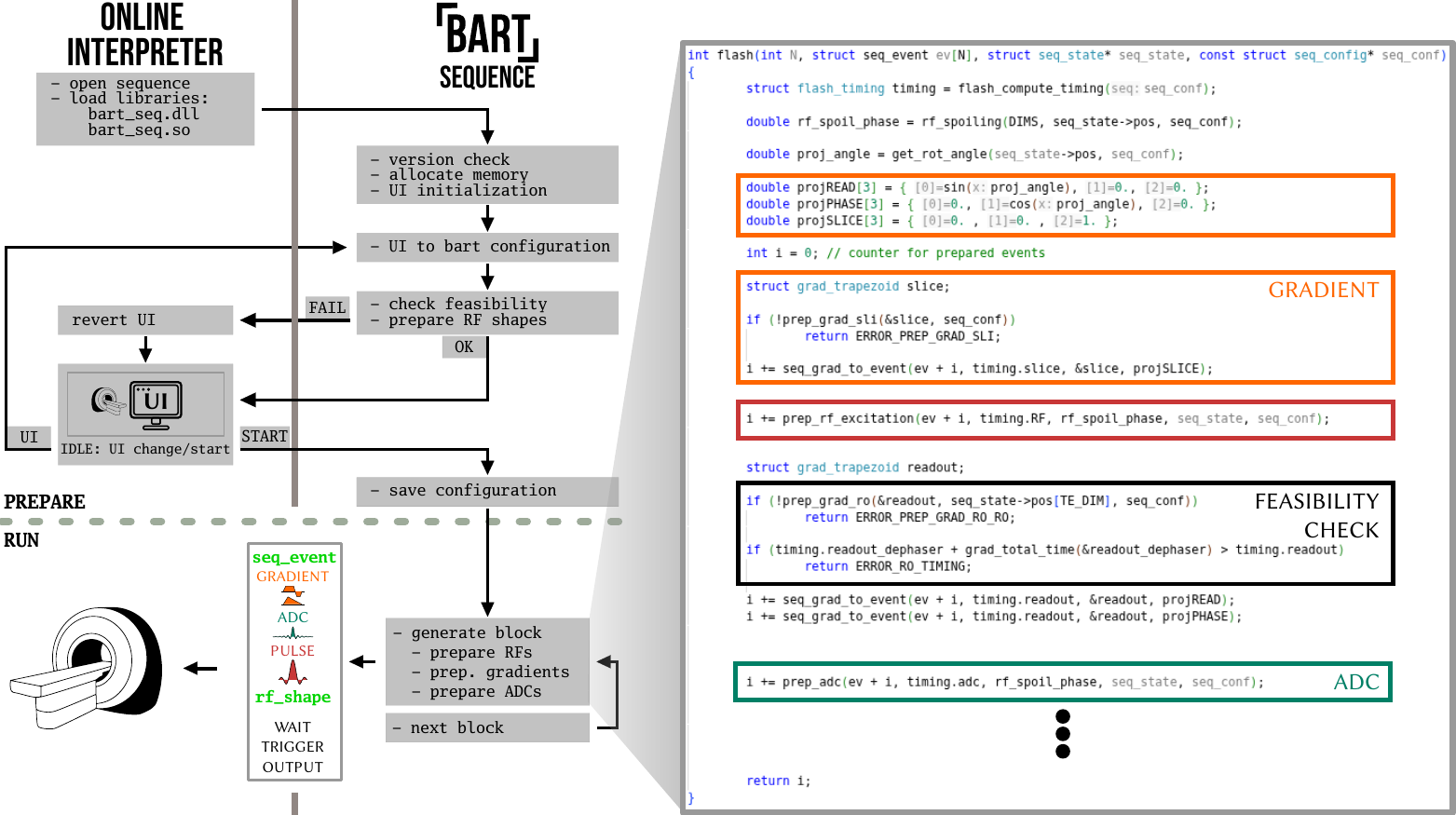}
	\caption{(A) Sequence preparation workflow: opening a sequence starts loading
		of specific BART library, followed by a version check, memory
		allocation and initialization of the UI. UI parameters are converted
		to the sequence configuration and reverted if the sequence is invalid.
		RF pulses are discretized and prepared in the vendor-specific format.
		When starting the acquisition, information of the configuration is
		written to a log file. When running the sequence, events are prepared 
		and interpreted block-wise.
		(B) Implementation of a gradient-echo imaging block including
		start times computation, spoiling phase and radial projection angle
		calculation. Event preparation of trapezoidal gradients with
		projection to logical axes as well as preparation of RF pulses and
		acquisition events.}
	\label{fig:dll}
\end{figure}

For online acquisition, we compile BART as a dynamic library and link
it with a vendor-specific driver sequence that can interpret the
generated sequence events. A flow chart of the interaction between the library 
and driver is shown in Figure \ref{fig:dll}.

The interface is designed in a generalized way to
completely disentangle the actual sequence logic from the driver
sequence to ensure the flexibility for further development
of new sequence types without modifying the driver sequence.
The complete interface to the vendor-dependent
driver sequence is described by four C headers: \textit{event.h, seq.h,
helpers.h, custom\_ui.h}. To ensure modularity,
internal data structures of the sequence are not exposed to the driver
sequence.
Hence, sequence libraries implementing new sequences can be linked to 
the driver sequence without any need to update this vendor-specific 
component. Compatibility of driver sequence and library is ensured by 
a mutual version check at run-time (\texttt{bart\_seq\_version\_check}) 
which is performed during an initial handshake. To avoid mistakes, a 
specific sequence library must also be paired to the driver sequence 
using a digital signature.

Vendor-specific user interface (UI) parameters are mapped to sequence parameters
using a generalized interface that also allows the definition of
custom parameters. For every parameter change, feasibility of the configuration 
set is checked. During execution, sequence events are computed just-in-time
and interpreted/converted to actual vendor-specific hardware
instructions for every sequence block.
The generated block is typically designed as a modular
functionality, e.g. a spoiled GRE imaging consisting of an excitation pulse,
gradients for k-space encoding and ADC for acquisition.

An exemplary implementation of an imaging block is shown in Figure \ref{fig:dll}.
Prior to event preparation, the start times are determined. The RF phase used
for excitation pulse and ADC and the radial projection angle are computed.
Gradients are prepared as trapezoids and a failed preparation returns an error,
feasible gradient amplitudes are projected onto corresponding axes.
Preparation of RF pulses include mapping to a shape and complex modulation
depending on the spoiling phase and the slice shift. Within a block, various
feasibility checks can be included. Analogous to the pulse, complex
modulation of the ADC phase depends on the spoiling phase and in-plane
FOV shift. In order to save the acquired data at the correct position, the
current looping positions are added to the ADC event.

For reproducibility of acquisitions, a log file containing a synthesized
command-line is automatically saved each time a sequence is started.
Also, one can extract all required data from the exported raw data. 
This information set combined with the actual implementation is
used to reproduce an identical sequence, which ensures that all
information that may be required for (model-based) reconstruction
is faithfully preserved.

\subsection{Offline Sequence Tool and Pulseq Export}

Timings and waveforms of a sequence can also be computed offline using the \texttt{bart seq}
command based on a high-level parameter set as shown in \ref{ssec:sub} and \ref{ssec:moba}.
Within this command, the sequence is interpreted as in the driver sequence.
Outputs of the command currently include gradient waveforms and k-space
positions, as well as k-space position, time, and phase of each ADC sample.
This allows extraction of the k-space trajectory for image reconstruction
and retrospective computation of FOV shifts. 

The events can also be saved in Pulseq format. Continuous gradients 
and pulse shapes are discretized on a raster.  ADC events include additional 
metadata information for labeling. Trigger events are described via the corresponding Pulseq
extension type. For more complicated sequences with multiple ADCs, we split blocks
during conversion to the Pulseq format. Data can then be acquired with vendor-specific Pulseq 
interpreters. The complete command-line used for sequence generation is also 
stored in the Pulseq file.

\section{Example Applications}

We demonstrate the framework using examples of increasing
complexity.
As a basic example, we demonstrate an online sequence for radial MRI with 
full support for in-plane FOV shifts. Here, the trajectory used
for reconstruction is derived from the sequence implementation.
To show feasibility of the framework for
the development of advanced computational MRI applications, we implemented
two quantitative MRI applications that combine radial acquisition with 
model-based reconstruction. 
The first quantitative method uses linearized subspace reconstruction
for $T_1$ mapping with Inversion-Recovery FLASH, while the second uses 
a non-linear model-based approach for joint mapping of $R_2^*$ and $B_0$
using multi-echo radial FLASH.

For each example, we validate
offline reproducibility by comparing the results from measurements 
performed online with results from a repeated measurement based
on an exported Pulseq file. The differences are compared to test-retest 
differences of the same acquisition strategy.

\subsection{Data Acquisition}

MRI experiments were performed on a 3~T Magnetom Vida (Siemens Healthineers,
Erlangen, Germany) with Pulseq v1.5.1 and the dynamically linked BART sequence. 

Phantom experiments were conducted with a commercial reference phantom
(CaliberMRI Inc, Boulder, CO USA) consisting of 14 $T_1$ spheres.
Phantom and brain studies were conducted
with a 20-channel head/neck coil and cardiac experiments were performed with
a combined thorax and spine coil with 26 channels. Brain and cardiac
scans were performed on a healthy volunteer (male, 29 years) with
written informed consent.

The implemented sequence utilizes a radial FLASH readout with rational
approximation of golden angles \cite{Scholand_Magn.Reson.Med._2025}.
Data was acquired with a base resolution of 256, FOV of 220x220~mm$^2$
(phantom/brain) or 256x256~mm$^2$ (cardiac), slice thickness 6~mm (3~mm brain), and
excitation flip angle of $\alpha~=~6^\circ$ ($20^\circ$ brain).
Measurements were acquired within $\approx 4~s$ resulting in a total
number of 1300 excitations for cardiac ($T_R~=~3.11~ms$),
1200 for phantom ($T_R~=~3.4~ms$), and 133 for multi-echo experiments ($T_R~=~31~ms$).

\subsection{FOV Shift for Non-Cartesian MRI}

An in-plane FOV shift can generally be realized for non-Cartesian MRI 
by applying a phase modulation
\cite{Magland_Magn.Reson.Med._2006}
$\Theta(t)~=2 \pi \boldsymbol{k}(t)\cdot \Delta\boldsymbol{r}~d\tau$
to the acquired data with $\boldsymbol{k}$ the k-space trajectory and $\Delta\boldsymbol{r}$ the FOV shift.
This modulation can be applied directly during Cartesian and radial
data during the acquisition by setting the frequency and phase used for demodulation.
For online acquisitions this is done in the driver sequence resulting
in intrinsically correct complex demodulation. For offline acquisition 
of non-Cartesian sequences with Pulseq, the
data is affected by phase inconsistencies
when the default method for FOV positioning is used. 
This can be corrected retrospectively by adding the actual FOV shifts to the
\texttt{bart seq}  command and using a script provided as part
of the BART toolbox.
\begin{align}
	&\texttt{\small bart seq -r 377 ... -s 0.0:0.1:0.0 adc \# shift of 100mm in y}~\nonumber\\
	&\texttt{\small \$BART\_TOOLBOX\_PATH/scripts/pulseq\_correction.sh -s 4.6E-6 adc corr}~\nonumber\\
	&\texttt{\small bart fmac ksp corr ksp\_corr}\nonumber
\end{align}

Pulseq interpreters modulate the ADC frequency and phase
as a function of the FOV shift. The calculation is not exactly correct,
which results in a delay
of $\approx~4.6~\mu s$ that was empirically
determined to optimize reconstruction quality and which corresponds
to about half the raster time.
To correct this, timing information and
complex phase computed by BART can be used. The complex phase is resampled
for each readout line while taking the delay into account and the k-space
data is then multiplied with a complex phase.

In the first experiment, differences of raw k-space data from isocenter placed
FOV were compared for online and Pulseq sequences.
For each method, two phantom measurements
with continuous radial readout of five frames with 377 spokes were performed.
Global phase changes are expected because temperature changes 
can lead to global $B_0$ drifts. An optimal global complex scaling factor 
was therefore applied to each subtrahend to remove this effect.
In a second experiment, images were reconstructed with NLINV
\cite{Uecker_Magn.Reson.Med._2008} and compared for isocenter and FOV-shifted
acquisitions for phantom and in vivo measurements. For Pulseq
acquisitions, phase inconsistencies due to the FOV shift were corrected
retrospectively by providing the FOV-shift as input
to the \texttt{bart seq} command and applying the phase modulation
to the data in a preprocessing step.

\subsection{$T_1$ Mapping with Inversion-Recovery Radial FLASH}
\label{ssec:sub}

The single-shot sequence for T1 mapping starts with a non-selective
adiabatic inversion pulse, followed by a continuous radial FLASH readout with
random RF phase spoiling \cite{Bernstein__2004,Frahm_Magn.Reson.Med._1986,Roeloffs_Magn.Reson.Med._2016}.
Cardiac measurements were conducted under breath-hold and the inversion pulse
was ECG-triggered to the early diastolic phase
\cite{Wang_Magn.Reson.Med._2018,Wang_J.Cardiovasc.Magn.Reson._2019}.
The cardiac sequence was generated with the following command:
\begin{align}
	\texttt{\small bart seq}
	&~\texttt{\small --trigger --trigger-delay 0.4 --IR\_NON  --slice\_thickness 6E-3}~\backslash\nonumber\\
	&~\texttt{\small --FOV 0.256 --BR 256 --dwell 4E-6  --rf\_duration 620E-6 --FA 6}~\backslash\nonumber\\
	&~\texttt{\small --BWTP 3.8 --flash --raga -r377 -t1350 --TE 1.9E-3 --TR 3.11E-3}~\backslash\nonumber\\
	&~\texttt{\small -R raga\_indices adc\_samples gradients moments ir\_cardiac.seq}\nonumber
\end{align}

We performed subspace reconstruction where the signal evolution is constrained
to a low dimensional linear subspace \cite{Petzschner_Magn.Reson.Med._2011,
huang_Magn.Reson.Med.2012,Zhao_Magn.Reson.Med._2015,Tamir_Magn.Reson.Med._2017,
Roeloffs_Magn.Reson.Med._2018,Wang_Philos.Trans.R.Soc.A._2021}. 
The Look-Locker model \cite{Look_Rev.Sci.Instrum._1970} is given by
\begin{align}
	&M(t,\boldsymbol{x_p}(r))~=~M_{ss}(r)~-~(M_{ss}(r)~+~M_{0}(r))
			~e^{-t/T_1^*(r)}\;, \label{eq:ll}\\
	&\texttt{\small bart signal -F -I -1 \$T1\_MIN:\$T1\_MAX:\$N1 -5 \$FA\_MIN:\$FA\_MAX:\$N2}~\backslash\nonumber\\
	&~~~~~~~~~~~~~~~~~~~\texttt{\small-r\$TR -n\$SPOKES --short-TR-LL-approx dicc}\nonumber
\end{align}
with the model parameters $\boldsymbol{x_p}(r)=(M_0, M_{ss}, T_1^*)^T$ where
$M_{ss}$/$M_0$ are the steady-state and equilibrium magnetization and $T_1^*$
the apparent longitudinal relaxation time. We computed a signal dictionary for
variable $T_1$ and nominal flip angles and performed an
PCA to extract the first four left-singular basis vectors
($B_c(t)$). The signal is approximated by a linear combination with
corresponding coefficient maps $\boldsymbol{x_c}(r)$:
\begin{align}
	&M(\boldsymbol{x_p}(r),t)~\approx~\sum_c^{N_c}~\boldsymbol{x_c}(r)~B_c(t)~.\\
	&\texttt{\small bart svd dicc - S V | bart extract 1 0 N$_C$ - basis}\nonumber
\end{align}

We precomputed coil sensitivities ($\mathcal{C}$) with subspace-NLINV
\cite{Blumenthal_Magn.Reson.Med._2024} and solved
the linear problem to estimate the coefficient maps.
\begin{align}
	&\texttt{\small bart nlinv -g -i12 -B basis -t traj ksp tmp coils}\nonumber\\
	&\hat{\boldsymbol{x_c}} = \arg \min_{\boldsymbol{x_c}}
	||\mathcal{P~F~C~D~B} ~\boldsymbol{x_c}~-~\boldsymbol{y}||_2^2
	+ \lambda \mathcal{R}(\boldsymbol{x_c})\\
	&\texttt{\small bart pics -geH -i250 -RW:3:64:3e-3 -B basis -t traj ksp coils coeffs}\nonumber
\end{align}
with the temporal basis $\mathcal{B}$, a temporal mask $\mathcal{D}$ selecting
diastolic data, the Fourier
transform $\mathcal{F}$, the sampling pattern $\mathcal{P}$ and $\mathcal{R}$
a regularization operator, e.g. joint $\ell_1$-wavelet.
The physical parameters were fitted pixel-wisely according to
\begin{align}
	&\hat{\boldsymbol{x_p}}~=~\arg~\min_{\boldsymbol{x_p}}~||\mathcal{B}^H~
	\mathcal{D~B}~\boldsymbol{x_c}~-~\mathcal{B}^H~\mathcal{D~M}
			(\boldsymbol{x_p})||_2^2~.\\
	&\texttt{\small bart mobafit -g -B basis -L --init 1:1:0.8 TI coeffs pars}\nonumber
\end{align}
Finally, the Look-Locker correction $T_1~=~M_0/M_{ss}~T_1^*~+~2~T_d$ with
a delay time $T_d$ (time between inversion and first acquired spoke) was
applied to obtain the $T_1$ map \cite{Deichmann_J.Magn.Reson._1992}.
\begin{align}
	&\texttt{\small bart looklocker -t0 -D\$T$_d$ pars t1map}\nonumber
\end{align}

\subsection{Model-based $R_2^*$ and $B_0$ Mapping with Multi-Echo Radial FLASH}
\label{ssec:moba}
The sequence starts with preparation excitations to achieve a steady-state
magnetization, followed by a four-second continuous radial readout. 15 echoes
with blip gradients between each readout were used to obtain a uniform k-space coverage.
The sequence was generated with the following command:
\begin{align}
	\texttt{\small bart seq}
	&~\texttt{\small --prep\_scans 100  --FOV 0.220 --slice\_thickness 3E-3 --flash}~\backslash\nonumber\\
	&~\texttt{\small --BR 256 --dwell 46E-7  --rf\_duration 13E-4 --FA 20 --BWTP 3.8}~\backslash\nonumber\\
	&~\texttt{\small --raga -e15 --tiny 26 -r133 --TE 2.55E-3 --TE\_delta 1.9E-3}~\backslash\nonumber\\
	&~\texttt{\small --TR 31E-3 -R raga\_ind adc\_samples gradients moments meco.seq}\nonumber
\end{align}

The signal evolution in the steady-state
for the $m$th echo can be described by
\begin{align}
	S_{\text{TE}_{m}}(\boldsymbol{x_p}) &=
	\Big[\rho_W + \rho_F \cdot z_{m}\Big]\cdot\exp\big(\text{TE}_{m}
	\cdot i2\pi f_{B_{0}}\big)
	\cdot \exp\big(- \text{TE}_{m} \cdot R_{2}^{*}\big)\,,
\label{eq:meco}
\end{align}
where the set of unknowns $\boldsymbol{x_p}$ contains $\rho_W$ and $\rho_F$
the signals for water and fat, respectively, $f_{B_{0}}$ represents the field
inhomogeneity and $R_{2}^{*}$ is the effective transversal relaxation rate.
\cite{Tan_Magn.Reson.Med._2019,Tan_IEEE_TMI_2023}
$z_{m}$ describes the 6-peak fat spectrum. \cite{Middleton_ISMRM_2009}
A nonlinear forward operator $F$ is constructed to map the unknowns
of Equation (\ref{eq:meco}) to the acquired multi-channel data $\boldsymbol{y}$ at
$\text{TE}_{m}$, i.e.,
\begin{equation}
	F: \boldsymbol{x}=(\boldsymbol{x_p},\boldsymbol{c})\mapsto\boldsymbol{y}=\mathcal{PF}(\boldsymbol{c}\odot S_{\text{TE}_{m}}(\boldsymbol{x_{p}}))~.
\label{eq:forward}
\end{equation}
Also here, $\mathcal{P}$ is the non-uniform sampling pattern and $\mathcal{F}$
the Fourier transform and $\mathcal{C}$ the coil sensitivities.
The estimation of unknown $\boldsymbol{x} = (\boldsymbol{x_{p}}, \boldsymbol{c})^{T}$ of Equation \ref{eq:forward}
is then formulated as the optimization problem solved
\begin{align}
	&\boldsymbol{\hat{x}} = \text{argmin}_{\boldsymbol{x}\in~D}
	\sum_{\text{TE}}\|\mathcal{PFC}\cdot S_{\text{TE}_{m}}(\boldsymbol{x}) - 
	Y_{\text{TE}_{m}}\|_{2}^{2} +
	R(\boldsymbol{x}).
\end{align}
\begin{align}
	\texttt{\small bart moba}
	&~\texttt{\small -i20 -g -D -m1 -R3 -o 1. -k --kfilter-2 -C500 -j7e-5}~\backslash\nonumber\\
	&~\texttt{\small --other echo=TE ... -t traj ksp TI reco sens}\nonumber
\end{align}
$D$ is a convex set for non-negativity of $R_2^*$ and $R(\cdot)$
is the regularization term for both, parameter maps and coil sensitivity maps.
In particular, we use Sobolev regularization on  the coil sensitivity maps
\cite{Uecker_Magn.Reson.Med._2008} and the field inhomogeneity $f_{B_{0}}$
map to enforce smoothness and a joint $\ell_{1}$-Wavelet
sparsity constraint \cite{Wang_Magn.Reson.Med._2018} on remaining parameters
to exploit sparsity and correlations between maps.

\newpage
\section{Results}

\begin{figure}
	\centering
	\includegraphics[width=\textwidth]{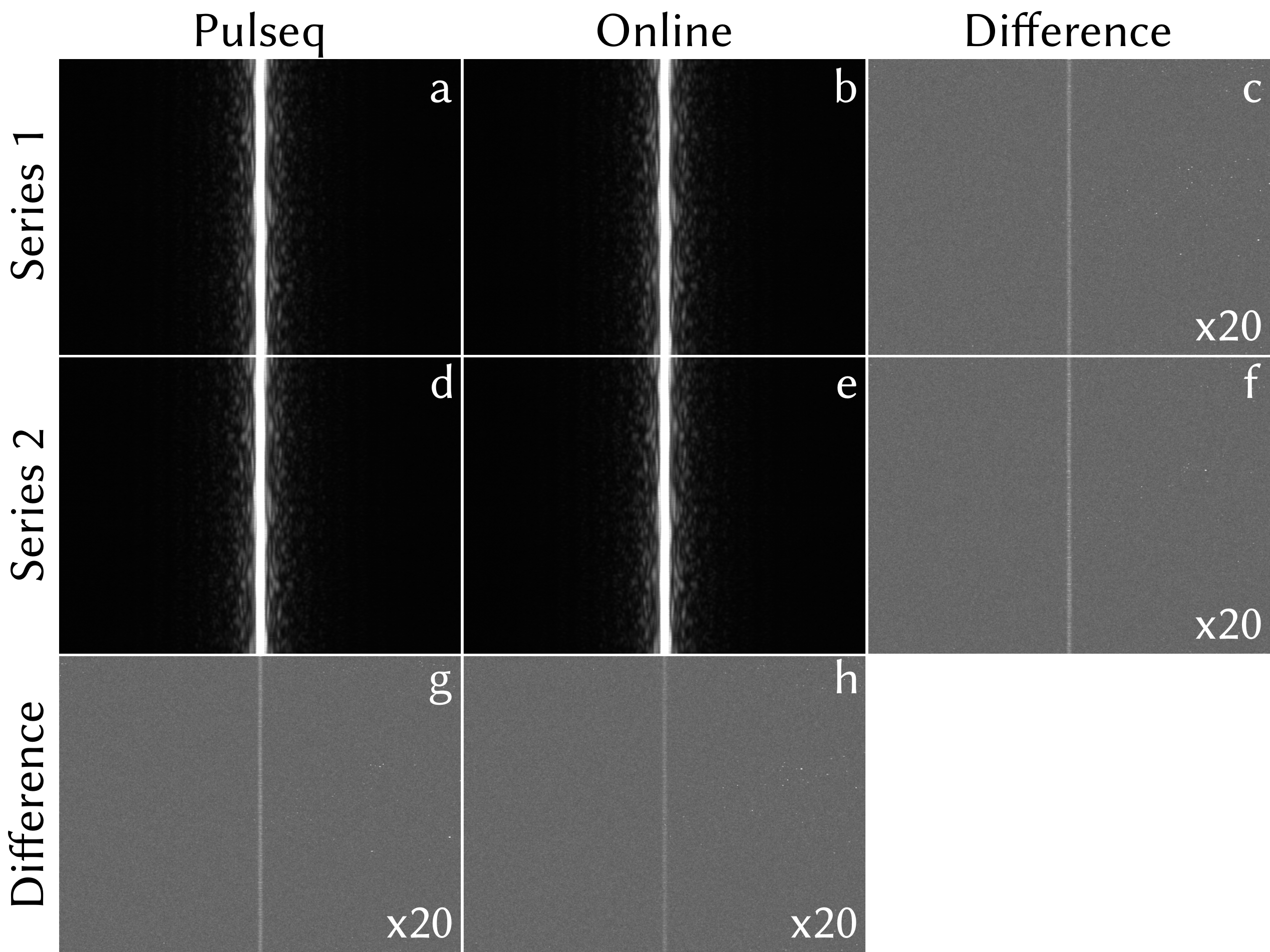}
	\caption{Test-retest experiments show only noise-like differences for
		both, Pulseq and online acquisitions  (g, h). Similarly, small
		differences can be observed between Pulseq and online 
		acquisition (c, f) with identical sequence parameters.}
	\label{fig:test_retest}
\end{figure}

Figure \ref{fig:test_retest} shows the differences between measurements
using online acquisition and using Pulseq. After correction with complex global
scaling factors, the difference of the k-space differences for both
are comparable to test-retest results, confirming offline reproducibility
of the scan.

\begin{figure}
	\centering
	\includegraphics[width=\textwidth]{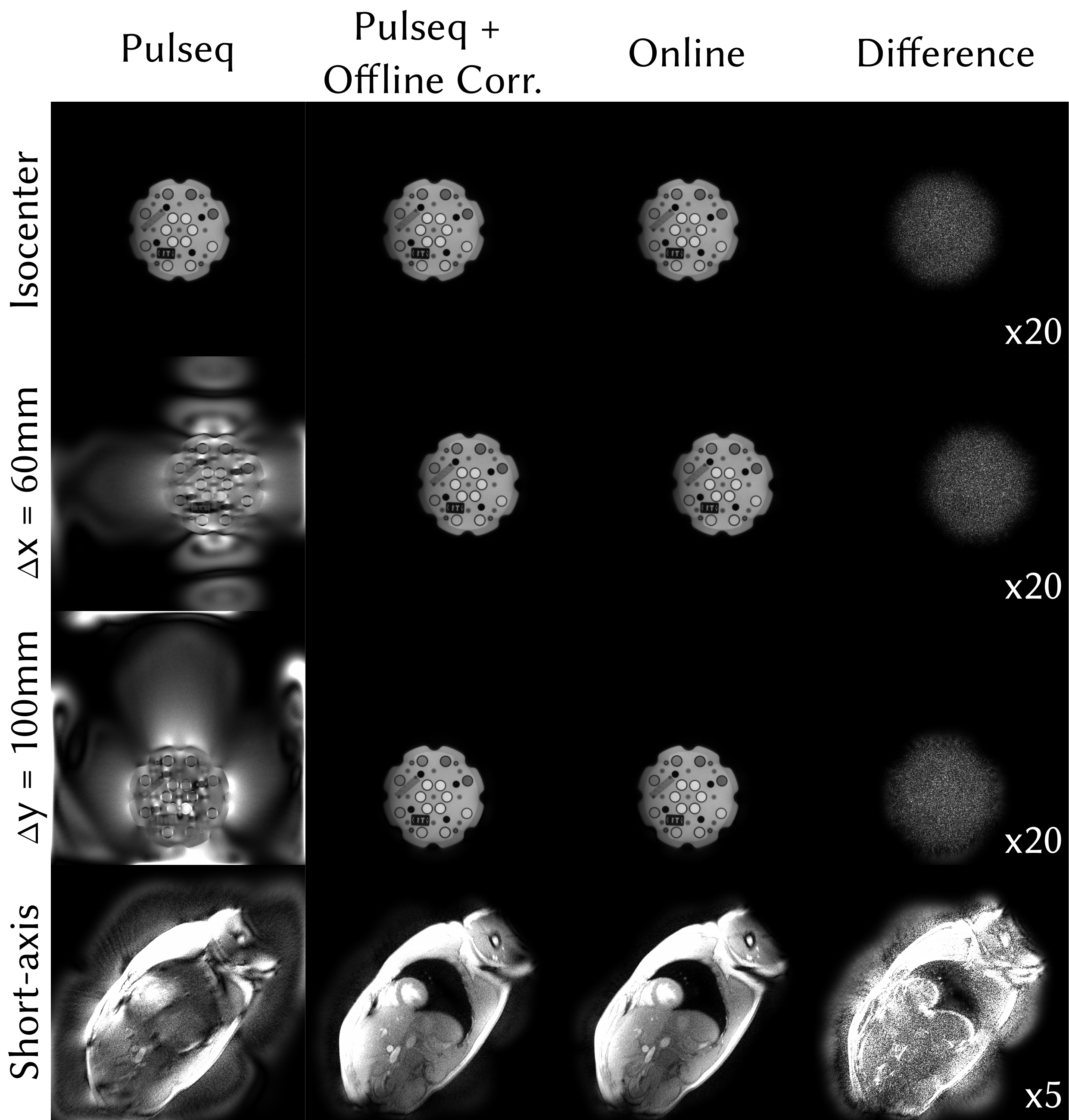}
	\caption{Online and offline Non-Cartesian acquisition with FOV shift.
		Pulseq acquisitions are affected by phase errors.
		After retrospective correction, only noise-like differences
		remain between online acquisition and Pulseq. This is 
		also confirmed for the cardiac measurement, although larger
		differences remain due to motion.}
	\label{fig:offcenter}
\end{figure}

In Figure \ref{fig:offcenter}, we compare images acquired
with radial FLASH again using both variants, Pulseq and online acquisition.
When acquiring data in the isocenter, only noise-like differences can be
seen between both methods for the phantom scan.  When using the
vendor-specific Pulseq interpreter with FOV shift, phase inconsistencies cause
artifacts. After retrospective correction, the comparison of FOV-shifted and corrected
Pulseq data to online acquisitions with inherently correct in-plane shifts
again shows only noise-like differences comparable to the isocenter experiment.
In a cardiac short-axis experiment, the same effect can be seen, although
larger differences between the two measurements can be observed due to
physiological changes between the two measurements, which is expected.

\begin{figure}
	\centering
	\includegraphics[width=\textwidth]{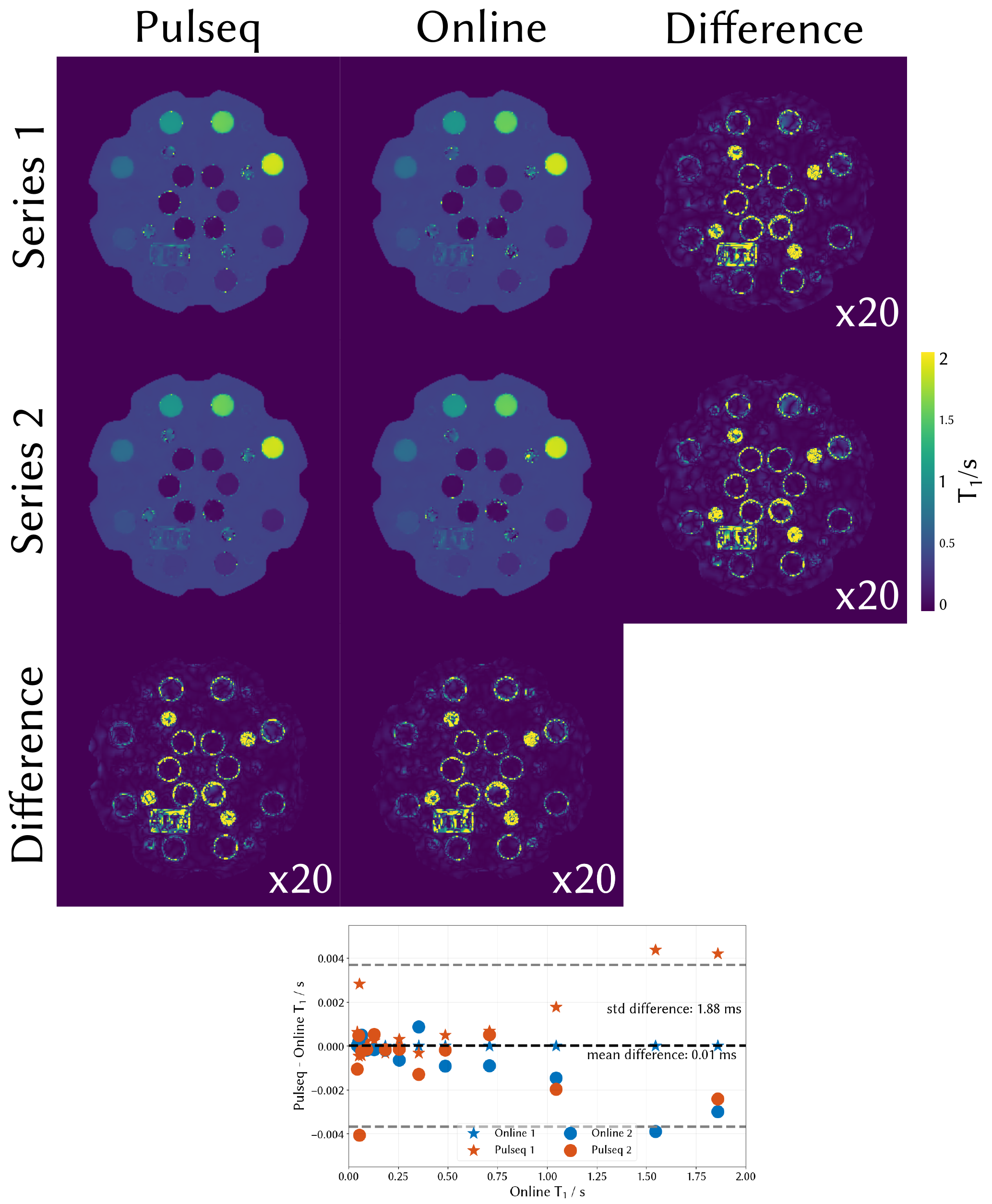}
	\caption{Comparison T1 mapping with  a single-shot IR-FLASH sequence with
		continuous radial readout and subspace-based linear reconstruction
		of the NIST phantom acquired with Pulseq and online acquisition in the
		isocenter. Excellent agreement can be found across
		ROIs within a wide range of T1 values. Again, differences between Pulseq
		and online acquisitions are similar as test-retest differences of the
		same acquisition strategy.}
	\label{fig:nist}
\end{figure}

T1 maps of the NIST phantom placed in the isocenter show only small visual differences
(Figure \ref{fig:nist}), for both, Pulseq compared to online acquisitions and
test-retest of the same acquisition strategy.
Evaluation of the 14 $T_1$ values (ranging from $42\:ms$ to $1.87\:s$)
results in a mean difference of $0.01\:ms$ (standard deviation $1.88\:ms$) between
the acquisitions.

\begin{figure}
	\centering
	\includegraphics[width=\textwidth]{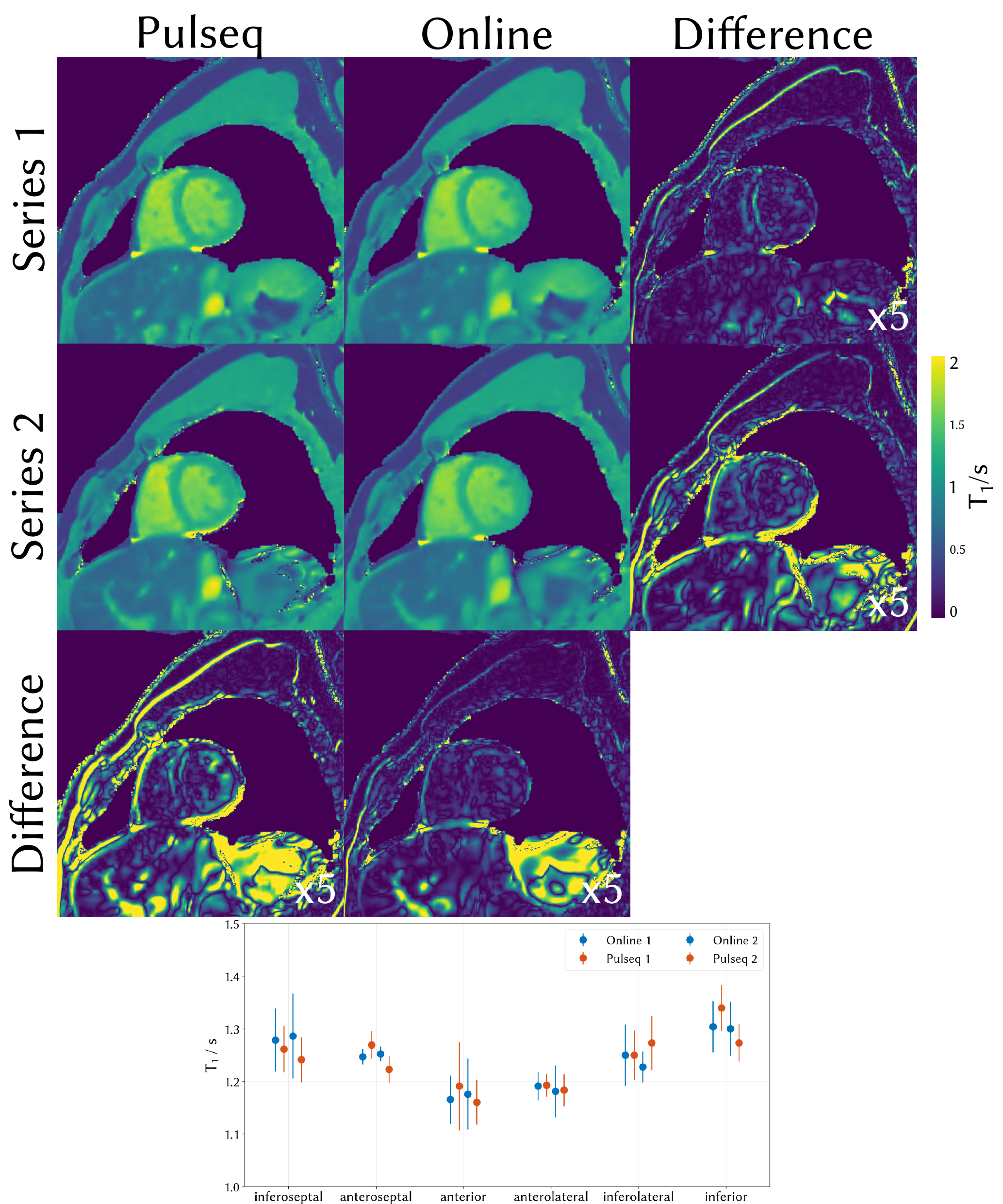}
	\caption{Cardiac T1 mapping with ECG-triggered radial IR-FLASH acquisition
		and linear subspace reconstruction utilizing Pulseq and online integration.
		After successful phase correction for the FOV-shifted Pulseq acquisition, 
		good agreement of T1 values can be achieved. Visually, this agreement
		is confirmed by similar difference maps for test-retest experiments with
		differences in both, comparison of Pulseq and online acquisitions and
		test-retest, mainly due to physiological motion.}
	\label{fig:cardiac}
\end{figure}

In Figure \ref{fig:cardiac} cardiac T1 maps of a Pulseq and online
acquisition in two consecutive series are presented, where phase
inconsistencies were successfully
corrected for the Pulseq datasets. Small residual visual differences are shown
in difference maps scaled by a factor of five. Comparison of the two acquisition strategies show
similar variability as test-retest with inter-measurement deviations mainly due
to physiological changes. A mean difference of $10\:ms$ (standard
deviation $20\:ms$) between the acquisitions is found across six standardized
myocardial ROIs \cite{AHA_Circulation_2002}.

 \begin{figure}
	\centering
	\includegraphics[height=0.8\textheight]{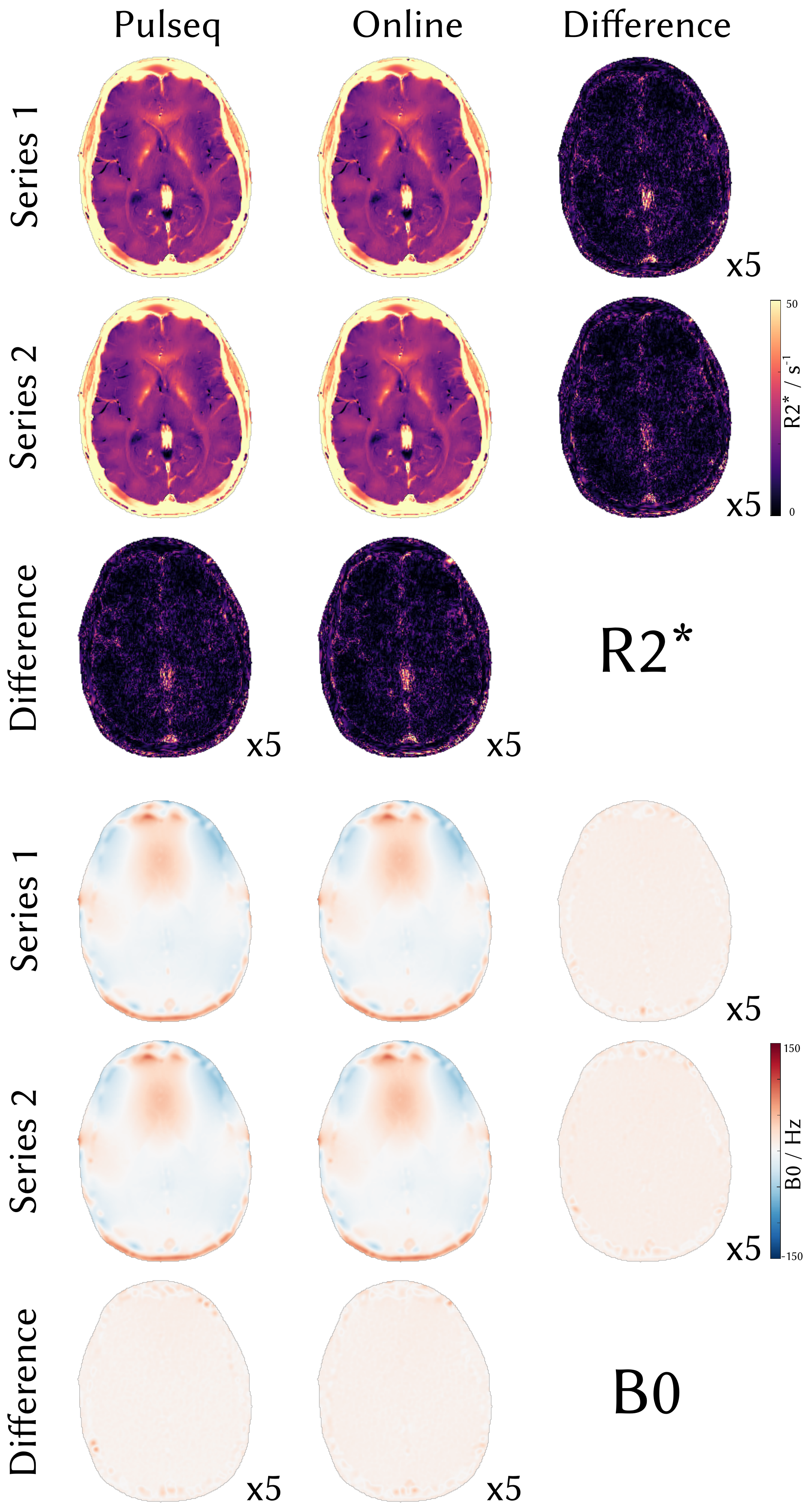}
	\caption{Radial multi-echo data acquisition (4 seconds) and model-based
		reconstruction for joint $R_2^{\star}$ and $B_0$ mapping.
		Good agreement between Pulseq and online acquisitions can be
		observed for both quantitative maps demonstrated by low differences
		across the whole brain. This is confirmed by similar differences between test-retest experiments of the same acquisition strategy.}
	\label{fig:meco}
\end{figure}

Model-based reconstructions of a FLASH sequence utilizing 15 echoes are
shown in Figure~\ref{fig:meco}. Good agreement can be found in the
parameter maps for both, $R_2^{\star}$ and $B_0$ maps, between Pulseq and 
online measurements. Again, similar deviations between acquisition methods are
observable as for test-retest experiments.

The comparisons of online and Pulseq acquisitions for both applications
show good agreement in the parameter maps. A successful correction for off-center
measurements is also demonstrated in the cardiac example. Similar $T_1$ maps are
obtained for both strategies and differences can be explained with physiological
changes, e.g. different breath-hold positions and a different number of spokes
in diastolic phase. Good agreement can also be found for the $R_2^*$ and $B_0$ maps
of the brain multi-echo acquisition where small differences can be observed in
the magnified difference maps. For both applications, the differences between Pulseq
and online acquisitions are comparable to test-retest differences of the same acquisition
strategy.

\newpage
\section{Discussion}

In this work, we extended BART, a widely used open-source toolbox for image
reconstruction, to a comprehensive end-to-end framework for computational
MRI that now also includes a sequence development framework. Furthermore,
we integrated BART as a library to a vendor-specific interpreter sequence,
which enables online parameter adjustment.
This driver is currently available for scanners from Siemens
Healthineers (Erlangen, Germany). In addition, we implemented an
offline tool that can recreate the same sequence and export it as a 
Pulseq file, suggesting compatibility with Pulseq drivers for various
MRI scanners and with simulation frameworks.
Newer versions of Pulseq interpreters also address the phase inconsistency
problem by including an implementation of the same FOV positioning algorithm
directly linked to the scanner's UI \cite{Shafiekhani__2025}.

An integrated framework for sequence development and reconstruction
offers many advantages as k-space trajectories, gradient and pulse
waveforms, and timing information can be recovered directly from the 
sequence and used in model-based reconstruction algorithms.
As examples for integrated computational MRI method where
sequence and model-based reconstruction are tightly integrated,
we show examples of highly-accelerated (4 seconds) $T_1$ mapping 
and joint $R_2^*$/$B_0$ mapping.  After online acquisition, an 
identical sequence can be recomputed, which ensures that all
information that may be required for model-based reconstruction
is faithfully preserved.

The integration into the development workflow of the BART toolbox
ensures long-term reproducibility and maintenance for sequences. 
Using a continuous integration pipeline, we  automatically run tests
when the software is changed to be able to identify and fix bugs early on. 
Tests are implemented on different levels, ranging from unit tests for
low-level functionality such as event preparation and block generation
to integration tests for the \texttt{bart seq} tool.

In this work, we focus on online acquisition and reproduction of
the same sequence for advanced quantitative MRI, which is shown for two applications.
As even more advanced applications, a radial arterial spin labeling
sequence and chemical exchange saturation transfer technique with
joint T1 mapping using dynamic transitions were recently presented
\cite{Buchegger_ISMRM_2026,Huemer_Magn.Reson.Med._2026,Huemer_ISMRM_2026}, demonstrating the
potential of this framework. A next step is to directly integrate
the framework with a Bloch-model based reconstruction \cite{Scholand_Magn.Reson.Med._2023}.

Although not further addressed in this work, online image reconstruction
can be implemented using the BART streaming framework as described recently
\cite{Schaten_Magn.Reson.Med._2026} and an example for non-Cartesian
MRI is distributed together with our vendor-specific driver sequence.
In order to enable online execution also on scanners from
other vendors, a corresponding driver sequence needs to be implemented.

\section{Conclusion}

An end-to-end open-source framework for sequence design and model-based
reconstruction based BART can be used to develop advanced qMRI techniques.
Online integration with a vendor-specific interpreter enables online
adjustment of sequence parameters, while Pulseq export allows
an exact offline reproduction of the sequence.

\FloatBarrier

\section*{Conflict of Interest}
The authors declare no competing interests.

\section*{Data Availability Statement}
In the spirit of reproducible research,
the code to reproduce the results of this paper
is available at \url{https://gitlab.tugraz.at/ibi/mrirecon/papers/bart-sequence}.
All reconstructions have been performed with BART \cite{bart__2026}.
BART is available at \url{https://codeberg.org/mrirecon/bart/}.
The data used in this study is available at Zenodo
\doi{10.5281/zenodo.21456517}.
The Pulseq interpreter can be obtained from the University of Freiburg.
The \textit{BART Driver Sequence} (BOOST: BART Online Open-Source Sequence
Toolbox) can be obtained from the Graz University of Technology via
Siemens Healthineers' C2P platform.

BART is continuously being developed and maintained since 2013. Functionality 
of the toolbox and reproducibility of publications is ensured by regular
automated testing and a public mailing list is available for support.
The Institute of Biomedical Imaging at  Graz University of Technology and
BART developers are committed to supporting the toolbox in the future.

\section*{Acknowledgement}
\printfunding

The authors thank the ISMRM Reproducible Research Study Group for
conducting a code review of the code (Version 0.1) supplied in the
Data Availability Statement.
The scope of the code review covered only the code’s ease of download,
quality of documentation, and ability to run, but did not consider
scientific accuracy or code efficiency.

\newpage
\printbibliography

\end{document}